# An inverse-designed nanophotonic interface for excitons in atomically thin materials

Ryan J. Gelly [1,†], Alexander D. White [3,†], Giovanni Scuri [1,2,3], Xing Liao [1,2], Geun Ho Ahn [3], Bingchen Deng [1,2], Kenji Watanabe [4,5], Takashi Taniguchi [4,5], Jelena Vučković [3,*], Hongkun Park [1,2,*]

[1]Department of Physics, and [2]Department of Chemistry & Chemical Biology, Harvard University, Cambridge, MA 02138, USA

[3]Department of Electrical Engineering, Stanford University, Stanford, CA 94305, USA

[4]Research Center for Functional Materials, and [5]International Center for Materials Nanoarchitectonics, National Institute for Materials Science, 1-1 Namiki, Tsukuba 305-0044, Japan

† These authors contributed equally to this work

*To whom correspondence should be addressed: jela@stanford.edu, hongkun_park@harvard.edu

# ABSTRACT

Efficient nanophotonic devices are essential for applications in quantum networking, optical information processing, sensing, and nonlinear optics. Extensive research efforts have focused on integrating two-dimensional (2D) materials into photonic structures, but this integration is often limited by size and material quality. Here, we use hexagonal boron nitride (hBN), a benchmark choice for encapsulating atomically thin materials, as a waveguiding layer while simultaneously improving the optical quality of the embedded films. When combined with photonic inverse design, it becomes a complete nanophotonic platform to interface with optically active 2D materials. Grating couplers and low-loss waveguides provide optical interfacing and routing, tunable cavities provide a large exciton-photon coupling to transition metal dichalcogenides (TMD) monolayers through Purcell enhancement, and metasurfaces enable the efficient detection of TMD dark excitons. This work paves the way for advanced 2D-material nanophotonic structures for classical and quantum nonlinear optics.

Keywords: 2D materials, nanophotonics, inverse design, integrated photonics, optical cavity

# MAIN TEXT

**Introduction**

Two-dimensional (2D) materials host a variety of classical and quantum light sources. Atomically thin direct-bandgap semiconductors like the transition metal dichalcogenide (TMD) monolayers possess tightly bound excitons[1] that recombine to emit light. They also host quantum defects that can serve as on-chip, scalable single-photon sources[2-5]. Optically active spin qubits in hexagonal boron nitride (hBN) are also promising candidates for quantum sensors and quantum information processing nodes[6].

Over the years, numerous efforts have been directed toward integrating 2D materials with SiN nanophotonic devices[7-10] and plasmonic nanocavities[11] to control their radiative properties. The typical scheme for coupling 2D light sources to a nanophotonic structure is to pre-fabricate a nanostructure and transfer the 2D material onto it[12]. This scheme, while effective, has several drawbacks: it does not straightforwardly integrate with fully encapsulated materials, alignment accuracy is limited by the transfer technique, and the 2D material is only evanescently coupled to the nanophotonic structure. It is now well established that the highest quality electrical[13] and optical[14,15] devices incorporating 2D materials come from samples encapsulated by hexagonal boron nitride (hBN). Encapsulation protects the optically active medium from environmental effects that are most prominent at the exposed surface. Therefore, recent efforts have focused on creating nanophotonic devices by directly etching the encapsulating hBN, which serves a dual purpose: maintaining the excellent optical quality of the atomically thin semiconductor, and providing a resonant photonic structure[16-21].

Here we introduce a new nanophotonics platform for hBN-encapsulated atomically thin materials. We first develop a fabrication technique to create clean etched photonic structures of hBN with feature sizes down to 20 nanometers. By combining this technique with photonics inverse-design, we demonstrate the key elements of a photonic interface – grating couplers for optical inputs and outputs, low loss waveguides for optical routing, and tunable optical resonators to enhance light matter interactions (Fig. 1a). We also demonstrate the capabilities of inverse-design to enable new optical functionalities by constructing a metasurface to brighten dark excitons.

**Design and fabrication of hBN nanophotonic structures**

We illustrate our approach using an encapsulated $MoSe_2$ monolayer as our model system. Figure 1b illustrates a schematic of a device defined using an $hBN/MoSe_2/hBN$ heterostructure on a $SiO_2/Si$ substrate: it consists of a partially etched top hBN layer that acts as a rib waveguide, a fully encapsulated $MoSe_2$ monolayer in the center, and a bottom hBN layer to complete the encapsulation. The $SiO_2$ layer thickness is chosen to be 2,000 nm to minimize loss in the absorptive Si layer. Figure 1c shows the simulated electric field profile for the transverse electric (TE) guided mode of the structure, with the $MoSe_2$ monolayer experiencing 93% of the maximal field amplitude. Such a good overlap between the optically active medium and the field maximum contrasts with the typical integration of TMD monolayers with SiN waveguides where the monolayer would be on the surface, rather than embedded. The respective thicknesses of 130 nm for the top hBN and 20 nm for the bottom hBN are chosen to best co-localize the $MoSe_2$ monolayer with the mode center while maintaining single mode operation and a higher index contrast for inverse design flexibility.

Two challenges stand in the way of translating the simulated device structure to a working device. First, high-quality hBN is not available as a wafer-scale (or even die-scale) thin film. While much effort is being directed toward this direction[22], as of now, the only source of clean hBN with the requisite thickness is via mechanical exfoliation, limiting the device footprint to approximately 100 μm on a side (Fig. S1). Meanwhile, typical photonic devices can have footprints of many hundreds of micrometers or millimeters. The second challenge is the nanofabrication of hBN. The dry etching of hBN traditionally used to generate devices for over a decade[23] is a notoriously dirty process, leading to a search for alternative methods, such as focused ion beam (FIB) milling[24].

We address these problems, first by inverse-designing compact nanophotonic structures subject to a hBN size constraint. We then optimize the etching of hBN to faithfully translate the design into a fabricated nanophotonic structure with < 100 nm feature sizes. The resulting structure can also contain TMD layers as optically active media.

To fabricate devices, we use a dry transfer method[12] to assemble a van der Waals heterostructure of hBN/MoSe$_2$/hBN on an Si/SiO$_2$ substrate. We then lithographically pattern and partially etch the top layer of hBN. In order to remove contamination produced during the etching process and enable the fabrication of small features, we include an acid washing step. While the lithographic and etching process is not novel, we are not aware of previous work that explores the compatibility of hBN with acids used in semiconductor wafer cleaning. The acid cleaning step uses a Piranha mixture (sulfuric acid and hydrogen peroxide) to first remove any organic contamination, and then uses the SC-2 mixture (hydrochloric acid and hydrogen peroxide) to remove charges embedded during the etching process that can negatively affect exciton optical properties. The result of this acid cleaning process is an etched hBN film that has low surface and sidewall roughness suitable for nanophotonics (here we achieve 0.9 dB/mm single mode waveguide propagation loss). Indeed, the atomic force microscopy image in Fig. 1d reveals the topography of an etched honeycomb pattern before and after acid washing. The minimum achievable feature size is 75 nm spatial periods and 20 nm widths (Fig. 1e). Notably, we do not completely etch-through the top hBN layer so that the TMD monolayer is not exposed to the acids during processing. Additional details regarding the nanofabrication process are available in the Methods section.

Equipped with a clean nanofabrication process applicable to hBN, we design and fabricate nanophotonic devices. The single-mode waveguide structures in Fig. 1b-c are straightforward to design and fabricate, but coupling into these structures requires some additional infrastructure. Edge couplers cannot be used since the flakes do not span the whole chip. Instead, we use grating couplers to couple from a free space objective into a waveguide mode. To minimize the footprint of these grating couplers, and thus pack as many devices onto a single hBN flake as possible, we utilize inverse-design software that we previously reported[25]. We optimize the coupling from the free space mode of our objective to the waveguide mode at a wavelength of $\lambda = 750$ nm subject to the constraints that the coupler has a footprint of 2 x 2 $\mu m^2$ and a minimum feature size of 90nm. For all simulations, we use readily available optical properties of hBN[26]. The inverse-designed structure and its predicted efficiency are shown in Fig. 2a. The maximum expected coupling efficiency for a 150 nm thick hBN coupler with a 2 x 2 $\mu m^2$ footprint is 12%.

**Characterization of hBN nanophotonic devices**

The grating coupler designs are lithographically patterned onto the hBN flake via the dry etching process outlined above and are used to probe the properties of the intervening waveguide in a device **D1**. Figure 2a (inset) presents an SEM image of the fabricated structure. We characterize the waveguide loss by generating two side-by-side waveguides in a single hBN flake of uniform thickness. A white light source is focused onto the first coupler while the light emitted from the second coupler is collected and sent to a spectrometer to measure transmittance through the structure (Fig. 2b). There is a 100 μm length difference between the two waveguides, so by measuring the ratio of the transmittances of the two waveguides we can determine the loss due to the waveguide alone, as losses due to the couplers, tapers, and bends will divide out. We find that the loss due to the waveguide is 9 dB/cm, which is sufficiently low for the small-footprint devices in hBN (additional details on how the waveguide loss is measured can be found in Fig. S3).

Having measured the waveguide loss, we turn our attention to the efficiency of the couplers themselves. We probe the transmission of a straight waveguide device and normalize by the source spectrum. The transmittance is plotted in Fig. 2c. The maximum transmittance is related to the coupler efficiency by $T = g_{in}\alpha_{wg}g_{out}$ where $\alpha_{wg}$ is the loss in the waveguide, while $g_{in}$ and $g_{out}$ are the coupler's input and output efficiencies, which are expected to be equal $g_{in}=g_{out}=g$ by reciprocity. Therefore, g has a value of $g = \sqrt{\frac{T}{\alpha_{wg}}}$. For this device with the measured maximum transmittance of 1.2%, the maximum g is estimated to be 11%, in good agreement with the simulated value of 12%. The shift in the wavelength of maximum efficiency compared to the simulation is likely due to infidelity of pattern transfer during fabrication. This can be ameliorated by fabricating an array of devices slightly scaled relative to each other.

To prove that these BN-based structures remain effective when the TMD monolayer is embedded, and to verify that the processing steps do not degrade the optical response of the encapsulated TMD, we also probe waveguides with grating couplers with embedded monolayers of MoSe2 (Fig. S4).

In addition to waveguides and couplers, nanophotonic resonators are critical to enhance the emission from excitons and their coupling to our photonic structures. Here we use inverse design to generate highly reflective mirrors out of waveguides, analogous to 1D photonic crystals. This process allows for the flexibility to design mirrors well matched to the intrinsic material loss and thus enables higher transmission than traditional photonic crystal resonators[27]. Figure 2d shows the generated structure and an SEM image after fabrication. Placing two of these mirrors face-to-face with a gap (Fig. 2e) forms a nanophotonic cavity where two dimensions of confinement are provided by the waveguide, and the third by the mirrors. The mirrors are designed to have a reflectance of 99.5% leads to a cavity (simulated in Fig. 2e) with a simulated finesse of 625 and expected quality factor of Q = 9,400 for a 5-μm-long cavity. Measuring the transmittance of this structure (device **D2**) reveals a cavity mode with $f$ = 735 nm (407.9 THz) and Q = 1415, as demonstrated in Fig. 2f.

**Coupling charge tunable TMD monolayers to hBN nanophotonic devices**

The main advantage of hBN as a nanophotonic material derives from its ability to be integrated with 2D light sources. To demonstrate such ability, we fabricate a device **D3** and characterize nanophotonic structures with embedded $MoSe_2$ monolayers. First a heterostructure of hBN (20 nm)/$MoSe_2$ monolayer/hBN (130 nm) is assembled and transferred to a Si/$SiO_2$ substrate (2,000 nm oxide). The heterostructure is etched into a waveguide with two couplers and with the waveguide having been fashioned into a cavity by incorporating two, one-dimensional photonic bandgap mirrors separated by 5 μm.

At T = 5K, a top-down photoluminescence (PL) spectrum from the **D3** cavity center confirms a high-quality optical response from the $MoSe_2$ monolayer, with the neutral exciton resonance at 752 nm and the charged exciton at 765 nm (Fig. 3a). When PL is collected via the grating couplers instead, three cavity modes are apparent, in addition to the neutral and charged exciton resonances. The cavity modes overlapping the tail of the charged exciton, has a quality factor of Q = 1,050 (Fig. 3a). Additional devices show quality factors of Q > 2,300 (Fig. S7). Fitting the charged exciton resonance with a Lorentzian and dividing the cavity mode maximum by the value of the fit at that frequency gives a Purcell enhancement of 15.

The cavity-exciton interaction can be tuned *in situ*. We make electrical contact to the $MoSe_2$ monolayer and use the doped Si substrate as a gate electrode to control the free carrier density in the monolayer (Fig. 3b). Figure 3c shows PL spectra as a function of gate voltage. The spectra exhibit five peaks: the neutral exciton ($X^0$), the charged exciton ($X^-$), and three cavity modes (M1, M2, and M3). Modes M2 and M3 have no dependence on the gate voltage, while the cavity mode M1 adjacent to the charged exciton shifts by 1.1 nm (greater than its linewidth, 0.7 nm), as highlighted in Fig. 3d. During the electrical tuning process, charged exciton radiative properties are actively modified due to both a reduction in Purcell enhancement and interaction with charges. The neutral exciton is only visible for positive gate voltages, suggesting that the MoSe$_2$ monolayer is intrinsic in this voltage range. Reflectance measurements also delineate the intrinsic and *p*-doped regimes based on the oscillator strength of neutral and charged excitons (Fig. S5). The presence of the charge-tunable MoSe$_2$ monolayer, therefore, enables the tuning of the cavity mode *in situ*.

We can understand this tuning mechanism in terms of the complex refractive index due to the excitonic resonance. In the vicinity of the exciton resonance $\omega_0$ the optical susceptibility of the MoSe$_2$ monolayer is given by: [28,29]

$$\chi(\omega) = -\frac{c}{\omega_0 d} \frac{\gamma_r}{\omega - \omega_0 + \frac{i\gamma_{nr}}{2}}$$

which in turn gives rise to the complex refractive index defined by $n(\omega) = \sqrt{1 + \chi(\omega)}$, where $\gamma_r$ and $\gamma_{nr}$ are the radiative and non-radiative decay rates for the exciton and $d$ is the monolayer thickness (0.6 nm). Calculating the complex index of the monolayer using the parameters for the charged exciton (Fig. S6a), we find that on resonance the index increases by $\Delta n = 20$ compared to the absence of the charged exciton resonance. And while the MoSe$_2$ monolayer represents only a small volume inside the cavity, such a large index shift modifies the effective index of the waveguide mode by 1.4% (Fig. S6b), leading to a predicted shift in the resonance by 1.1 nm, in excellent agreement with the observed shift.

**Observing dark excitons in the far-field**

So far, we have used inverse design has been to reduce the footprint and increase the efficiency of classical nanophotonic structures, but it can also be used to enable functionalities that are not possible with traditional design techniques. Here we use inverse design to optimize a metasurface that couples z-polarized dipole radiation into free space to measure dark excitons in TMDs.

Dark excitons have a transition dipole moment that is out-of-plane polarized (z-polarized, transverse magnetic (TM))[30]. Consequently, the dipole radiation from a bare dark exciton radiates preferentially into the plane, and exactly zero power is radiated into the solid angle perfectly normal to the plane. Emission from dark excitons can be collected by rotating the sample with respect to the optical axis[31], applying large in-plane magnetic fields[32], using near-field collection techniques[33], or by coupling it to TM modes via surface plasmon polaritons[34] or waveguides[30]. The z-polarized dark exciton transition only has zero radiated power exactly along the z-axis. At large angles to the z-axis, there is finite emission into the far-field (Fig. 4b) that can be captured by large numerical aperture (NA) objectives, but even with high NA objectives the collection efficiency is quite low.

To increase this collection efficiency, we optimize an inverse-designed metasurface made from the encapsulating hBN that preferentially scatters in-plane propagating light into the acceptance angle of the objective. This strategy is illustrated in Figure 4a. Figure 4b shows our design structure and its effect on the far-field distribution. We fabricate a device **D4** consisting of a 20 nm bottom hBN, a $WSe_2$ monolayer, and a 130 nm top hBN with the metasurface etched into the top hBN (Fig. 4c). We switched from a $MoSe_2$ monolayer to a $WSe_2$ monolayer in **D4** because in $WSe_2$ dark excitons have lower energies than their bright counterparts, unlike in $MoSe_2$. Probing the integrated PL from the metasurface coupled monolayer (Fig. 4d) shows a series of local maxima from the metasurface unit cell's centers. In addition to the metasurface brightening the integrated PL, the dark exciton is now the brightest feature in the PL spectrum (Fig. 4e). Compared to the much-diminished $X^D$ feature in a control spot (metasurface absent), the signal is amplified by 3.5x. This amplification is consistent across this device (Fig. S8) and between devices (Fig. S9).

**Discussion and Conclusion**

Armed with a fabrication process that produces high precision and low loss photonic structures, we demonstrate the key building blocks for interfacing TMDs and other 2D materials with nanophononics – grating couplers, low loss waveguides, and tunable cavities. Using inverse design, we show that additional functionality on top of traditional photonics can be built into the same system.

High-quality hBN nanophotonic devices should enable scientific and engineering advancements along several fronts, both in the realm of 2D materials and beyond. First, high-Q and small-mode volume nanocavities in hBN can interface with excitons in TMD monolayers to generate strongly coupled exciton polaritons. The tight confinement to the nanocavity will further exaggerate exciton-exciton interactions, leading to nonlinear optical phenomena based on these materials[35-37]. Moreover, nanocavities tuned to Rydberg excitons in these materials will be able to probe Rydberg exciton interactions[35], thus providing a promising platform for realizing Rydberg blockade and for studying quantum nonlinear optics in the solid state[38]. In addition, beam-shaping metasurfaces might be devised using inverse design to generate gate-switchable spatial-light-modulators based dark excitons (Fig. S10). Applications to van der Waals materials operating at longer wavelengths, such as in the infrared, could also be embedded into the hBN photonic platform with larger feature sizes that could take advantage of other fabrication techniques[39].

The hBN nanocavities enable the study of spin qubits in hBN, such as the boron vacancy center. These qubits exhibit weak optical transitions, making it difficult to observe single or few emitters as opposed to large ensembles. Enhancing emitter-photon coupling via nanophotonic cavities will pave the way for single-qubit control, enabling applications in quantum sensing and quantum information processing.

Finally, the fabrication technique presented here may also allow the *post hoc* modification of modular nanophotonic structures. The hallmark of van der Waals materials such as hBN is their lack of dangling bonds. Therefore, we may be able to generate both active and passive nanophotonic devices in hBN where a particular component can be swapped into and out of a device *post hoc* using pick-and-place dry transfer techniques.

# FIGURES

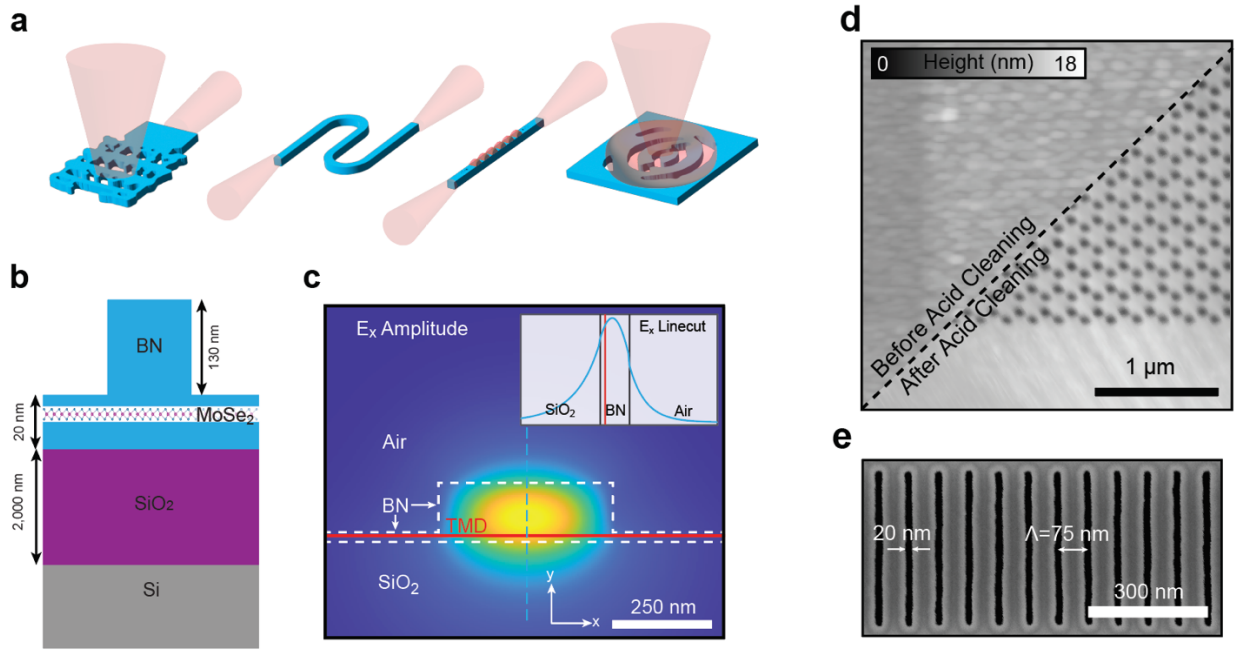

**Figure 1 – Device schematic, and fabrication process and results.**

(a) Schematic overview of the photonic platform based on hBN encapsulation. Key devices demonstrated include (from left to right): grating coupler, waveguide, cavity, and dark exciton metasurface. (b) A cross section of the proposed device structure where a $MoSe_2$ monolayer is encapsulated between two films of hBN (blue) on a $Si/SiO_2$ substrate (gray/purple) and then the top hBN film is etched into a waveguide. (c) Simulation of the guided mode profile for the dimensions given in (b) and at a wavelength of 750 nm. Inset: a linecut of the $E_x$ field component at the dashed blue line in the main panel. (d) The efficacy of the Piranha acid wash step is clear when the same region is compared pre- and post-wash. I Small feature sizes and spacings are possible by this process.

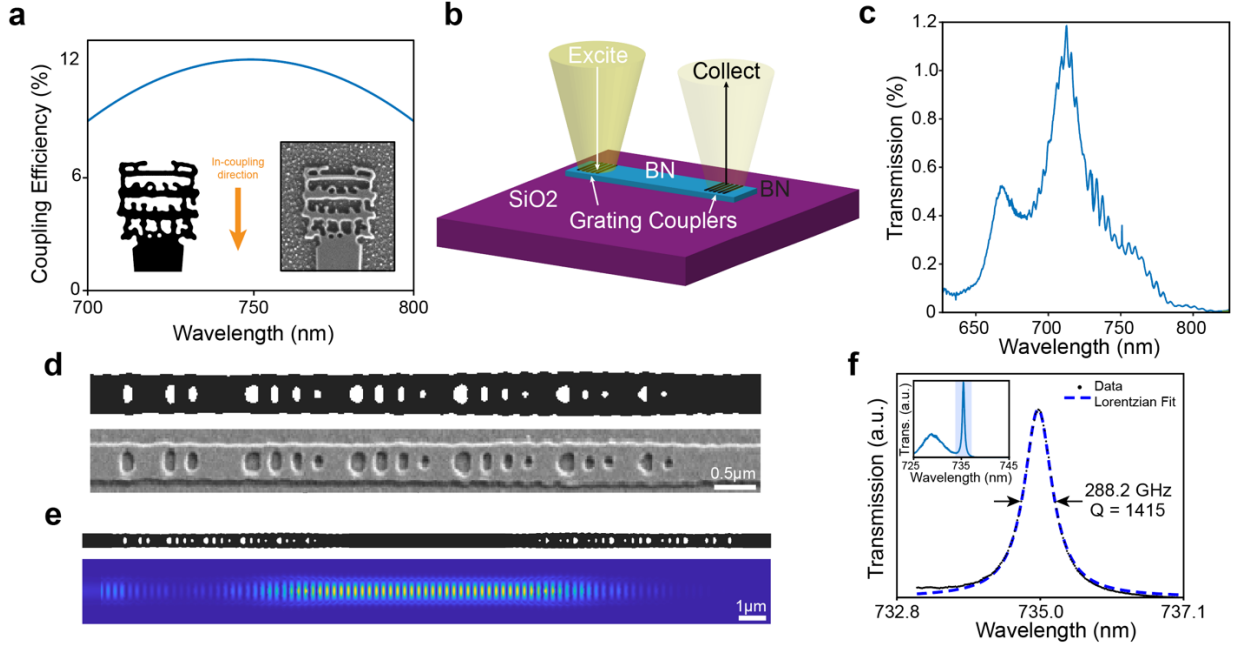

**Figure 2 – Characterization of hBN nanophotonic components.**

a) An inverse-designed grating coupler structure for launching a TE waveguide mode and its simulated efficiency as a function of wavelength. Inset: the designed structure and an SEM image of the fabricated device. (b) A depiction of the transmission spectrum measurement scheme. (c) The measured system throughput, which consists of two grating couplers, corresponding to a single-grating peak coupling efficiency of 11%. (d) The designed structure and an SEM image of a fabricated inverse-designed waveguide mirror for reflecting light at 750 nm. (e) A nanophotonic cavity is defined by two adjacent inverse designed mirrors from (d), and the cavity mode field profile is simulated. (f) The transmittance spectrum of this inverse-designed cavity.

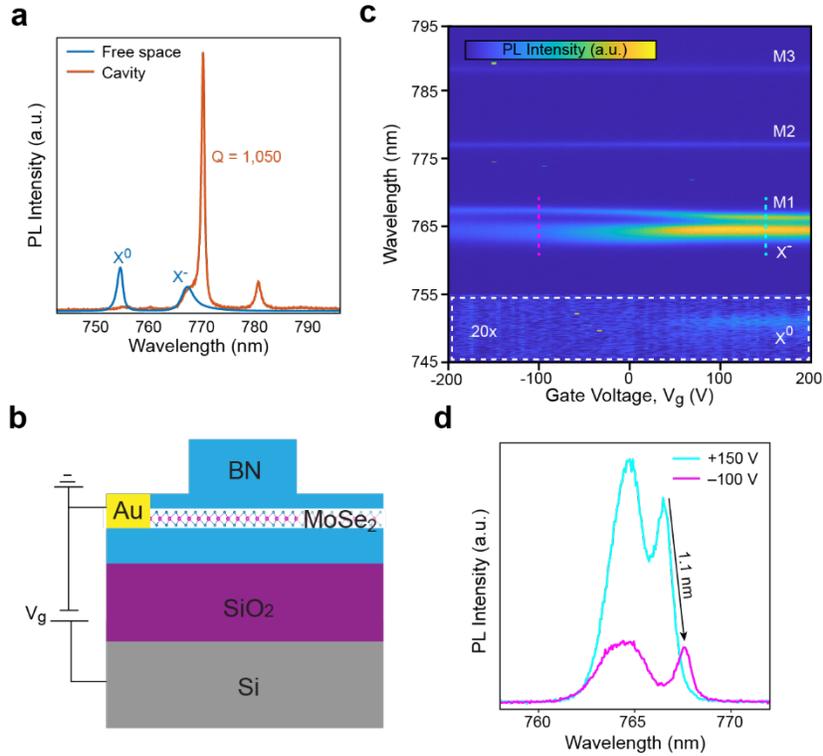

**Figure 3 – Gate control over cavity mode due to effective index change.**

(a) After integrating a MoSe$_2$ monolayer into the cavity introduced in Figure 2, Purcell enhancement of the charged exciton (X⁻) shoulder is observed. (b) The device is thermally cycled and an electrical contact is made to the MoSe2 monolayer and the Si substrate is used as a gate to control the carrier density. (c) The spectrum as a function of gate voltage shows a shift in the cavity mode (M1), charged exciton (X⁻) and neutral exciton (X⁰) as the carrier density is varied. Two other cavity modes (M1 and M2) are also present. The dashed region is amplified by 20x to see the neutral exciton feature. (d) Spectral linecuts of X⁻ and cavity mode M1 at gate voltages –100 V and +150 V show a 1.1 nm shift in M1.

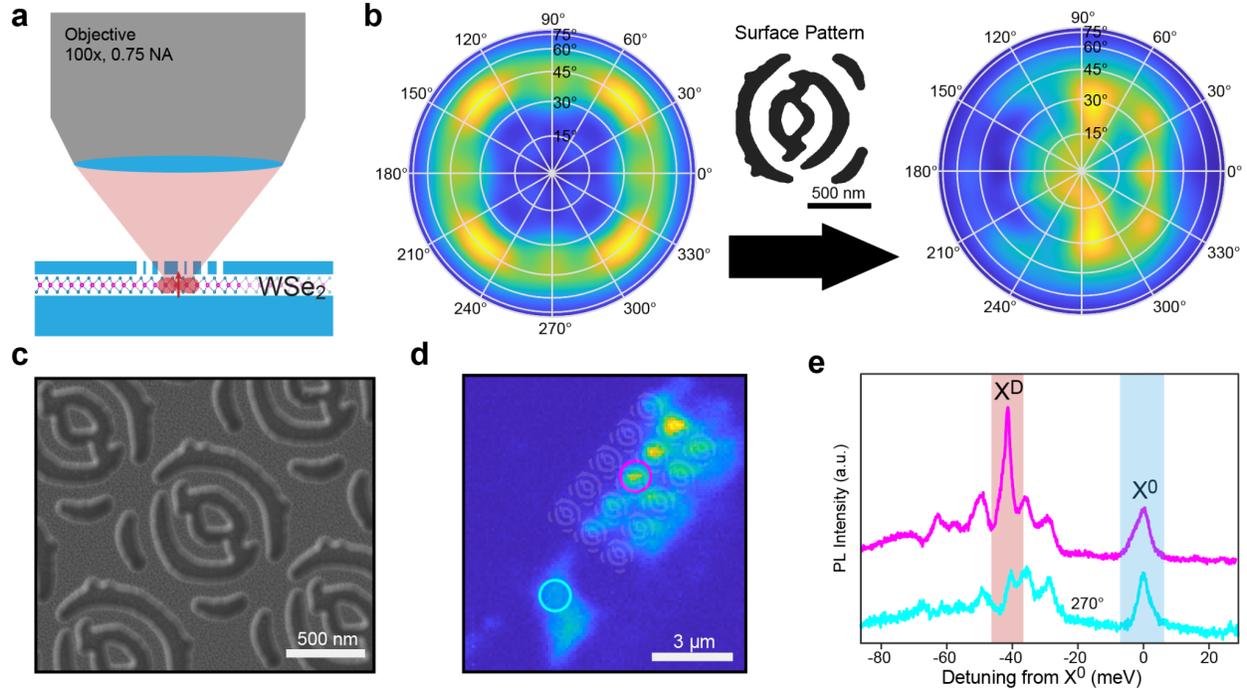

**Figure 4 – Observing dark excitons in the far field by an inverse designed metasurface.** (a) An outline of the scheme in which the top encapsulating hBN surrounding a WSe$_2$ monolayer is nanopatterned to scatter TM modes out into the acceptance angle of the objective. (b) Left: the far-field distribution of light from a z-polarized dipole embedded in 130 nm of hBN. Center: An inverse-designed surface structure is etched into the hBN. Right: the far-field distribution of light from a z-polarized dipole underneath the etched surface, which is more weighted toward the vertical axis. (c) SEM image of the fabricated surface. (d) A spatial map of PL (spectrally integrated) that shows enhanced emission at the centers of the metasurface unit cells. The etch mask design is superimposed on the PL map in semi-transparent Iy. (e) PL spectra from the two indicated regions in (d) where the metasurface region shows a prominent $X^D$ feature 40 meV below the $X^0$ resonance, consistent with the dark exciton based on previous literature. This is the brightest feature in the spectrum, while the unpatterned region has a low intensity $X^D$ feature, corresponding to the emission normally collected at high angles by the high-NA objective.

# METHODS

*Sample Fabrication*

Fabrication flow is illustrated in Fig S2. WSe$_2$ & MoSe$_2$ monolayers (HQ Graphene) and multilayer hBN flakes were mechanically exfoliated (Scotch Magic Tape) onto silicon substrates with a 285 nm silicon oxide layer. Monolayers of WSe$_2$ and MoSe$_2$ were identified by their contrast under an optical microscope and verified by their PL spectra. hBN layers of the appropriate thickness were identified by their color compared to a lookup table. Heterostructures were fabricated by a viscoelastic dry transfer method: a polycarbonate film on a PDMS stamp pick up successive flakes before dropping the stack and polycarbonate film onto a chip with bonding pads and markers. The polycarbonate was removed by washes in chloroform, acetone, and isopropanol. Next, electron-beam lithography (Elionix ELS-F125) was used to define the negative of the photonic structures in PMMA resist (PMMA 950 C2, spin coated at 5000 rpm and pre-baked at 180 °C) with a current of 300 pA and dose of 1600 µC/cm$^2$. The PMMA is developed in a 4 °C bath of 3:1 isopropanol to water mixture, followed by an isopropanol dip and nitrogen blow dry. The hBN exposed by lithography is etched by reactive ion etching (STS ICP RIE) using a 1:2:8 mixture of Ar to O$_2$ to CHF$_3$ at 10 mTorr and coil and platen powers both equal to 30 W. The etch rate is determined part way through by removing the sample and checking with atomic force microscopy (Oxford Asylum Instruments) before re-loading and finishing the etch at the determined rate. After etching the remaining PMMA is stripped in 80 °C NMP (Remover PG, Kayaku Advanced Materials) and then acid washed in 3:1 H$_2$SO$_4$:H$_2$O$_2$ Piranha mixture for 1 min followed by a 1:1:7 HCl:H$_2$O$_2$:H$_2$O SC-2 mixture, also for one minute. The sample is then rinsed in water and dipped in isopropanol before blow drying with nitrogen. The sample is finally mounted onto a cryogenic stage and Au wire bonds are made to the pre-fabricated bonding pads.

*Optical Measurements*

All optical measurements are performed in a home-built confocal microscope with an 0.75 NA, 100x objective lens at the sample. PL mapping was done by scanning a 532 nm, above band gap laser across the sample with galvo mirrors, while the PL is collected by avalanche photodetectors (APD). Optical spectra are collected by a Princeton Instruments spectrometer with blazed gratings with 300, 1200, or 1800 lines/mm. Transmission and reflection measurements utilize a halogen lamp or a supercontinuum laser (SuperK with Varia, NKT Photonics) as a white light source. Gate-dependent measurements were made with sourcemeters (Keithley 2400).

## SUPPORTING INFORMATION

Supporting Information: Additional information about the fabrication flow, experimental details, reflectance measurements and reproducibility across multiple devices.

## AUTHOR CONTRIBUTIONS

R.J.G., A.D.W., G.S., J.V., and H.P. conceived this project. R.J.G. fabricated the samples. A.D.W. and G.H.A. designed and simulated the structures. R.J.G., designed and performed the measurements with assistance from G.S., X.L., and B.D.. R.J.G. analyzed the data with help from A.D.W. T.T. and K.W. grew and provided the hexagonal boron nitride bulk crystals. R.J.G., A.D.W., G.S. wrote the manuscript with extensive input from all authors. H.P. and J.V. supervised the project.

## ACKNOWLEDGEMENTS

All fabrication was performed at the Center for Nanoscale Systems (CNS), a member of the National Nanotechnology Coordinated Infrastructure Network (NNCI), which is supported by the National Science Foundation under NSF award 1541959. We acknowledge support from the DOE (DE-SC0020115), ONR-MURI (N00014-20-1-2450), and Partnership for Quantum Networking from Amazon Web Services (A50791).

## COMPETING INTERESTS

The authors declare no competing interests.